\documentstyle[prd,amssymb,aps,epsfig]{revtex}
\def\bi{\bibitem}
\begin{document}
\baselineskip 0.5cm
\newcommand{\be}{\begin{equation}}
\newcommand{\ee}{\end{equation}}
\newcommand{\beq}{\begin{eqnarray}}
\newcommand{\eeq}{\end{eqnarray}}
\newcommand{\bear}{\begin{array}}
\newcommand{\ear}{\end{array}}
\newcommand{\D}{\displaystyle}
\newcommand{\bec}{\begin{center}}
\newcommand{\eec}{\end{center}}
\newcommand{\bed}{\begin{displaymath}}
\newcommand{\eed}{\end{displaymath}}
\title{The CWKB particle production and classical condensate in de Sitter spacetime.}    
\author{
S.Biswas$^{*a),b)}$, I. Chowdhury$^{**a)}$\\ 
a) Dept. of Physics, University of Kalyani, West Bengal,\\
India, Pin.-741235 \\
b) IUCAA, Post bag 4, Ganeshkhind, Pune 411 007, India \\
$*$  email: sbiswas@klyuniv.ernet.in\\
$**$ email: indrani@kyuniv.ernet.in 
, indrani\_chowdhury@indiatimes.com}

%\date{}
\maketitle
\begin{abstract}
The complex time WKB approximation is an effective tool in studying particle production in curved spacetime. We use it in this work to understand the formation of classical condensate in expanding de Sitter spacetime. The CWKB leads to the emergence of thermal spectrum that depends crucially on horizons (as in de Sitter spacetime) or observer dependent horizons ( as in Rindler spacetime). A connection is sought between the horizon and the formation of classical condensate. We concentrate on de Sitter spacetime and study the cosmological perturbation of $k=0$ mode with various values of $m/H_0$. We find that for a minimally coupled free scalar field for $m^2/H_0^2<2$, the one-mode occupation number grows more than unity soon after the physical wavelength of the mode crosses the Hubble radius and soon after diverges as $N(t)\sim O(1)[\lambda_{phys}(t)/{H_0^{-1}}]^{2\sqrt{\nu^2-1/4}}$, where $\nu\equiv (9/4 -m^2/{H_0^2})^{1/2}$. The results substantiates the previous works in this direction. We also find the correct oscillation and behaviour of $N(z)$ at small $z$ from a single expression using CWKB approximation for various values of $m/H_0$. We also discuss decoherence in relation to the formation of classical condensate. We also find that the squeezed state formalism and CWKB method give identical results.      
\end{abstract}

{\section {Introduction}} An important aspect of cosmological perturbation is that they are of quantum origin but eventually they become classical as they are (or source) responsible for classical density perturbation. A general trend to approach this problem is to consider the interaction between `subsystem' and `environment' such that the subsystem decoheres during evolution. The quantum fluctuations that originate in the early universe thereby undergo quantum to classical transition being stretched to astronomical size due to inflation. The works in this direction will be found in \cite{ms:ptp,rb:mpl,wu:prd,aa:hep,ec:prd,blh:osu,lpg:prd,vmuk:pr,bard:prd,cald:cqg,lpg1:prd,lidsey:rmp,ste:pl,habib:prd}. The decoherence due to particle production is another approach in which it is argued that the particle production during evolution decoheres the quantum fluctuations. Here we need not require to introduce interaction between the subsystem and the environment. Calzetta and Hu \cite{ech:prd} studied decoherence of a mean field by its quantum fluctuations, and Calzetta and Gonorazky \cite{cg:prd} considered fluctuations in a model with nonlinear coupling. In  \cite{mm:prd} decoherence is studied through particle production. The initial works in this direction are due to  Guth and Pi \cite{gp:prd}, Lyth \cite{ly:prd}, Polarski and Starobinsky \cite{ps:cqg}, and Lesgourges et al \cite{lps:np}.
\par
Let us try to understand how decoherence is obtained through particle production. Guth and Pi showed that at late times the wavefunction of long-wavelength modes behave as classical Gaussian probability distribution whereas Lyth pointed out that the wavepacket of massless modes in de Sitter space does not spread and this property is attributed to classicalization of fluctuation. In \cite{ps:cqg,lps:np} one looks at the free fields and the classicalization of the modes occur due to particular property of de Sitter spacetime. In this approach the decoherence emerges without invoking any interaction which is deemed essential in usual decoherence studies. One easiest way to study classical nature of a quantum variable is to look at the condition under which it becomes classical. For example, the canonical coordinate will behave like a classical variable if it approximately commutes with the canonically conjugate momentum. In expanding spacetime we are to choose a variable whose emergence is due solely to quantum behaviour of the fluctuation and in someway is related to classical density perturbation. One such quantity is particle occupation number. Assuming that vacuum fluctuations (that are supposed to act as seeds of density perturbation)in de Sitter spacetime during evolution produce particles, as is usually the case in expanding spacetime, we can search for decoherence looking at how this occupation number behaves at late times. If the number of produced particle is enormously large, we can argue that the situation is ripe enough for the formation of classical condensate. In other words decoherence is accomplished through particle production. This approach is sometimes known as `decoherence without decoherence'. In de Sitter spacetime the classical nature of fluctuation is usually `assumed, or argued for in passing, by invoking Gibbons-Hawking radiation, or the horizons in de Sitter space'. It is worthy to mention that a system with zero dissipation may well be in a pure quantum mechanical state, even though the occupation number is large. For this reason one may have some apathy to the idea of decoherence without decoherence. However the present attempt may be continued in the manner of \cite{boy:atp} to the standard picture of decoherence in which the large fluctuations plus the inflaton zero mode assemble into a new effective field. This effective field behaves classically and it is this object that rolls down. In this sense the study of formation of classical condensate might be a viable attempt to study decoherence. 

We, in this work, will follow \cite{mm:prd} but will adopt the CWKB (complex trajectory WKB) method of particle production. The reasons for doing so will be clarified in the text. We consider a massive field in de Sitter background and study the conditions under which the vacuum fluctuations in this spacetime evolve into classical fluctuations. We also discuss how the results obtained in the present work can be related to the standard method of studying decoherence through squeezed state formalism.

\par 
 To start with we discuss in section II the field theoretic approach to calculate the number $N_k$ of particles produced in a $k=0$ expanding de Sitter spacetime with a Hubble parameter $H_0$ for various values of mass of a free, scalar field. In section III we obtain the CWKB extension of $N_k$ that we adopt for numerical calculation. In section IV we discuss the numerical results obtained through CWKB analysis of section III. In section V we end up with a conclusion where we discuss the results obtained in the present working relation to squeeze state approach \cite{aa:hep,gjkks:ber}. 
\\
\\
\section{De Sitter Expansion}
Let $\phi$ denotes a scalar field of mass $m$, and $S(\eta)$ the scale factor for the Robertson-Walker spacetime with conformal line element. For the de Sitter case $S(\eta)=-\frac{1}{H_0\eta}$. The Lagrangian in a conformal spacetime reads
 (minimal coupling)
\be
\sqrt{-g}L=\frac{1}{2}S^2\eta^{\mu\nu}(\partial_\mu\phi)(\partial_\nu\phi)-\frac{1}{2}S^4m^2\phi^2
\ee
with the metric, $g_{\mu\nu}=S^2\eta_{\mu\nu},\,\,\eta_{\mu\nu}=\rm diag(1,-1,-1,-1)$. The field equation now reads 
\be
\frac{1}{S^2}\eta^{\mu\nu}\partial_\mu(S^2\partial_\nu\phi)+S^2m^2\phi=0
\ee
We now decompose the real scalar field into the modes as
\be
\phi(\vec{x},\eta)=\frac{1}{S}\int\frac{d^3k}{(2\pi)^3}e^{-i\vec{k}.\vec{x}}[u_{\vec{k}}(\eta)a_{\vec{k}}+u^*_{-\vec{k}}(\eta)a^\dagger_{\vec{k}}]
\ee
The mode function now satisfies the equation 
\be
\left[\partial_\eta^2+k^2+S^2m^2-\frac{1}{S^2}\frac{d^2S}{d\eta^2}\right]u_{\vec{k}}(\eta)=0
\ee
and the Wronskian
\be
u_{\vec{k}}(\eta)\partial_\eta u^*_{\vec{k}}(\eta)-u^*_{\vec{k}}(\eta)\partial_\eta u_{\vec{k}}(\eta)=i
\ee
For de Sitter spacetime, we get
\be
\left[\partial_\eta ^2+k^2+(\frac{m^2}{H_0^2}-2)\frac{1}{\eta^2}\right]u_{\vec{k}}(\eta)=0
\ee
The Hamiltonian corresponding to (1) is
\be
H=\int d^3x\frac{1}{2}\{S^2\left[(\partial_\eta\phi)^2+(\bigtriangledown \phi)^2\right]+S^4 m^2\phi^2\}
\ee
which in terms of creation and annihilation operators and the mode functions reads
\be
H=\frac{1}{2}\int\frac{d^3k}{(2\pi)^3}\{E_{\vec{k}}(\eta)(a_{\vec{k}}a^\dagger_{\vec{k}}+a^\dagger_{\vec{k}}a_{\vec{k}})+F_{\vec{k}}(\eta)a_{\vec{k}}a_{-\vec{k}}+F^*_{\vec{k}}(\eta)a^\dagger_{-\vec{k}}a^\dagger_{\vec{k}}\}
\ee
where in de Sitter time we find using eqn (4)
\beq
E_{\vec{k}}(\eta)&=&\left (\frac{1}{\eta}+\partial_\eta \right) u_{\vec{k}}(\eta)\left (\frac{1}{\eta}+\partial_\eta\right) u^*_{\vec{k}}(\eta)+\omega_k^2(\eta)u_{\vec{k}}(\eta)u^*_{\vec{k}}(\eta)\nonumber\\
F_{\vec{k}}(\eta)&=&\left (\frac{1}{\eta}+\partial_\eta\right)u_{\vec{k}}(\eta)\left (\frac{1}{\eta}+\partial_\eta\right)u_{-\vec{k}}(\eta)+\omega_k^2(\eta)u_{\vec{k}}(\eta)u_{-\vec{k}}(\eta)
\eeq
with 
\beq
\omega_k^2(\eta)& = & \vec{k}\,^2+\frac{m^2}{H_0^2}\frac{1}{\eta^2}\\
E_{\vec k}^2-\vert F_{\vec k}\vert^2 &=& \omega_k^2(\eta)
\eeq
The conformal time Heisenberg operators $a_{\vec k}, a^\dagger_{\vec k}$ in (8) can be written equivalently in terms of the $\eta$ dependent creation and annihilation operators $b_{\vec{k}}(\eta),\,\,b^\dagger_{\vec{k}}(\eta)$ that makes the Hamiltonian diagonal. Introducing the mode function $v_{\vec{k}}(\eta)$ with the field decomposition
\be
\phi(\vec{x},\eta)=\frac{1}{S}\int\frac{d^3k}{(2\pi)^3}e^{-i\vec{k}.\vec{x}}\left[v_{\vec{k}}(\eta)b_{\vec{k}}+v^*_{-\vec{k}}(\eta)b^\dagger_{\vec{k}}\right]
\ee
we get for the Hamiltonian
\be
H=\frac{1}{2}\int\frac{d^3k}{(2\pi)^3}\{E^\prime_{\vec{k}}(\eta)(b_{\vec{k}}b^\dagger_{\vec{k}}+b^\dagger_{\vec{k}}b_{\vec{k}})+F^\prime_{\vec{k}}(\eta)b_{\vec{k}}b_{-\vec{k}}+F^{\prime *}_{\vec{k}}(\eta)b^\dagger_{-\vec{k}}b^\dagger_{\vec{k}}\}
\ee
For $E^\prime_{\vec{k}}(\eta),\,\,F^\prime_{\vec{k}}(\eta)$ we get similar expressions as in (9) with $u_{\vec{k}}$ replaced by $v_{\vec{k}}$. The time dependent Bogolubov transformations that diagonalize the Hamiltonian i.e., $F_{\vec k}^\prime(\eta)=0$ now satisfy the relations 
\beq
u_{\vec{k}}(\eta) & = & \alpha_k(\eta)v_{\vec{k}}(\eta)+\beta_k(\eta)v^*_{-\vec{k}}(\eta)\nonumber\\
u^*_{-\vec{k}}(\eta) & = & \beta^*_k(\eta)v_{\vec{k}}(\eta)+\alpha^*_k(\eta)v^*_{-\vec{k}}(\eta)
\eeq 
and
\beq
b_{\vec{k}}(\eta)&=&\alpha_k(\eta)a_{\vec{k}}+\beta^*_k(\eta)a^\dagger_{-\vec{k}}\nonumber\\
b^\dagger_{-\vec{k}}(\eta)&=&\beta_k(\eta)a_{\vec{k}}+\alpha^*_k(\eta)a^\dagger_{-\vec{k}}
\eeq
with $\vert \alpha_k(\eta)\vert ^2-\vert \beta_k(\eta)\vert^2=1$. The reverse transformation of (14) and (15) can now easily be found out. One can easily check that the new creation and annihilation operators satisfy the standard algebra, and they act on a time dependent vacuum as
\be
b_{\vec{k}}(\eta)\vert 0>_\eta=0,\,\,\,\,b^\dagger_{\vec{k}}(\eta)\vert 0>_\eta=\vert \vec{k}>_\eta
\ee
For diagonalization, we see from (13) that, the mode function $v_{\vec{k}}(\eta)$ has to satisfy the condition
\be
F^\prime_{\vec{k}}(\eta)=0
\ee
Using (9) with $u_k$ replaced by $v_k$ we get from the above relation (17)
\be
\left(\frac{1}{\eta}+\partial_\eta\right)v_{\vec{k}}=-i\omega_kv_{\vec{k}}
\ee
Using this result in the expression of $E^\prime_{\vec{k}}(\eta)$ we get
\be
E^\prime_{\vec{k}}(\eta)=2\;\omega_k^2(\eta)\vert v_{\vec{k}}\vert^2
\ee
Using (8) and (13) we now get
\be
<0\vert 2b^\dagger_{\vec{k}}(\eta)b_{\vec{k}}(\eta)+1\vert 0 >=\frac{E_{\vec{k}}(\eta)}{E^\prime_{\vec{k}}(\eta)}
\ee
Hence
\be
N_k(\eta)=\frac{1}{2}\frac{E_{\vec{k}}(\eta)}{E^\prime_{\vec{k}}(\eta)}-\frac{1}{2}
\ee 
where
\be
N_k(\eta)=<b^\dagger_{\vec{k}}b_{\vec{k}}>=\vert \beta_k(\eta)\vert^2
\ee
is the number of particles referred to the initial Fock vacuum of the modes $u_{\vec{k}}$.  The result (20) can also be deduced from $F_{\vec k}^\prime(\eta)\equiv F_{\vec k}(\eta)(\alpha_k^*(\eta))^2-2E_k(\eta)\alpha_k^*(\eta)\beta_k(\eta)+F_k^*(\eta)(\beta_{k}(\eta))^2=0$. This gives $\alpha_k(\eta)=\beta_k^*(\frac{E_{\vec k}+\omega_k}{F_{\vec k^*}})$. Using $\vert \alpha_k\vert^2-\vert\beta_k\vert^2=1$, we get back the result (20). Further using $F_k^\prime(\eta)=0$ and $E_k^\prime(\eta)=\omega_k(\eta)$, we will find $\vert v_{\vec k}(\eta)\vert^2=(2\omega_k(\eta))^{-1}$ and hence (19) gives $E_k^\prime(\eta)=\omega_k(\eta)$. It should be pointed out that the conditions $F_k^\prime(\eta)=0,\,\,E_k^\prime(\eta)=\omega_k(\eta)$ do not determine the phase of $v_k(\eta)$. Using (9) and (18) we now get
\beq
N_k(\eta)&=&\frac{1}{4}\frac{(\frac{1}{\eta}+\partial_\eta)u_{\vec{k}}(\frac{1}{\eta}+\partial_\eta)u^*_{\vec{k}}+\omega^2_k(\eta)u_{\vec k}u_{\vec k}^*}{\omega_k^2(\eta)\vert v_{\vec{k}}\vert^2}-\frac{1}{2}\nonumber\\
&=& \frac{1}{4}\frac{\omega_k^2(\eta)\vert u_{\vec{k}}\vert^2+\vert u^\prime_{\vec{k}}\vert^2+\frac{1}{\eta}(\partial_\eta u_{\vec{k}}u^*_{\vec{k}}+\partial_\eta u^*_{\vec{k}}u_{\vec{k}})+\frac{1}{\eta^2}\vert u_{\vec{k}}\vert^2}{\omega_k^2(\eta)\vert v_{\vec{k}}\vert^2}-\frac{1}{2}
\eeq
So far the calculation is exact. Now to calculate $N_k(\eta)$ we are to find $u_k$ from (6) and $v_k$ from (18). We note that in (23) $\omega_k$ is always real and positive but in the mode eqn (6), we have the oscillator like equation
\be
u_k\,^{\prime\prime}+\tilde{\omega}_k^2u_k=0
\ee
where
\be
\tilde{\omega}_k^2=\omega_k^2-\frac{2}{\eta^2}=k^2+\frac{\alpha-2}{\eta^2}
\ee
with $\alpha=m^2/H_0^2$. Thus for $\alpha>2$, $\tilde{\omega}_k$ is everywhere real but for $\alpha<2$, $\tilde{\omega}_k^2$ turns out imaginary when $\eta\rightarrow 0$. In the literature most of the workers deal with real frequency but we will shortly find that the emergence of complex frequency has interesting features when one calculates $N_k$ through CWKB technique keeping the content of (23) unaltered. In our earlier works \cite{bis1:pmn,bis2:cqg,bis3:grg,bi:grg,sap:grg} we studied particle production through CWKB trajectory analysis with successful results.We adopt the same approach here. In Miji\'c's work, in defining the annihilation and creation operators as in standard harmonic oscillator prescription, he used $|\omega_k|$ instead of $\omega_k$. This allows in $dN_k/{d\eta}$ to have a part proportional to $\tilde{\omega}_k^2-|\tilde{\omega}_k|^2$. Some argue this to be an improper way to define the operators. In this work we eliminate this objection. Had one started with the standard harmonic oscillator like definition for $a$ and $a^\dagger$, one would not get the part that we just mentioned. We have tried to eliminate all these objections and explicitly show the role of complex trajectory through the use of WKB mode solutions. Moreover, in \cite{mm:prd} Miji\'c considered superposition of two WKB like solutions to get $u_2$ in evaluating $N_k$ at $z\rightarrow 0$ in contrary to the essence contained in (23). We give a clear physical arguments of this type of superposition in terms CWKB trajectory.
  
{\section{Complex trajectory and WKB approximation}}
The use of complex WKB trajectory to calculate particle production is well known \cite{bis1:pmn,bis2:cqg,bis3:grg,bi:grg,sap:grg,r23:dk}. The method has been used to calculate particle production in de sitter spacetime \cite{bis2:cqg} obtaining correct de Sitter temperature $kT=H_0/{2\pi}$. Subsequently the method has been used to study boson and fermion production using oscillating inflaton \cite{bcm:grg} with remarkable success that also settled some variant results in the literature. The reader may go through \cite{bis3:grg} to understand the basics of CWKB. In the present work we will quote some results of CWKB relevant to the context of the present paper. Before we pass to the CWKB calculation let us discuss Miji\'c's result in this regard. Considering (24) as a oscillator equation with time dependent frequency, and introducing
\beq
a_k&=& (-)i\sqrt{\frac{|\tilde{\omega}_k|}{2}}u_k+\frac{1}{2|\tilde{\omega}_k|}p_k\\
a_k^\dagger&=& (+)i\sqrt{\frac{|\tilde{\omega}_k|}{2}}u_k+\frac{1}{2|\tilde{\omega}_k|}p_k  
\eeq
and with the number operator $N_k=<a_k^\dagger\,a_k> $ it is shown that
\be
\frac{dN_k}{d\eta}=\frac{i}{2}\frac{\tilde\omega_k^2-|\tilde\omega_k|^2}{|\tilde\omega_k|}(a_k^{\dagger\,2}-a_k^2)-\frac{1}{2}\frac{|\tilde\omega_k|^\prime}{|\tilde\omega_k|}(a_k^{\dagger\,2}+a_k^2)
\ee

In obtaining (28) one has to use  $p_k\,^\prime=u_k^{\prime\prime}=-\tilde\omega_k^2\,u_k$ and replace $u_k$ and $p_k$ using (26) and (27) in terms of $a_k$ and $a_k^\dagger$. 
\par
 Now it is evident from the expression (28) that we have two sources of particle production. The first term contributes when the frequency is imaginary and vanishes for real frequency. The second term contributes for all frequencies. Most of the authors consider the second term in evaluating the particle production while considering the evolution of {\it in} vacuum to {\it out} vacuum through Bogolubov transformation technique between {\it in} and {\it out} vacuum. With (omitting the subscript $k$ on $u$)
\be
u_{1,2}(\eta)=\frac{1}{\sqrt{2\tilde\omega_k(\eta)}}e^{\pm\,i\int\,\tilde\omega_k(\eta)d\eta}
\ee 
the number of particles produced is obtained as
\be
N_k(\eta)=|\beta_k|^2=\frac{1}{2|\tilde\omega_k(\eta)|}|u_2^\prime|^2+\frac{|\tilde\omega_k(\eta)|}{2}|u_2(\eta)|^2-\frac{1}{2}
\ee 
Referred to (26) and (27), as noted earlier,  some have apathy in writing down the creation and annihilation operators in terms of $|\omega_k|$ for obvious reasons. Secondly, the solution $u_2$ is obtained in terms of Hankel function so that in the small $\eta$ approximation, one has to  put some restriction on $\alpha $ values. Miji\'c obtained large particle production when $\alpha<2$ and oscillating $N_k$ for $\alpha>2$. Leaving aside some minor technical points, we clarify the first point through field theoretic calculation as in (23). We then use WKB modes to calculate $u_2$ without using the exact solution of temporal equation (24). This is the advantage of CWKB over any other method. Moreover the definition of vacuum in curved spacetime is very delicate. The WKB like mode solution serves as a good template for defining the vacuum in curved spacetime. As we require only $u_2$ in calculating $N_k$, henceforth we call this solution as $u_k$ in the following.
\par
We now take the WKB modes as
\be
u_k(\eta)=\frac{1}{\sqrt{2\tilde{\omega}_k(\eta)}}e^{-i\int\,\tilde{\omega_k}(\eta)\,d\eta}  
\ee
as solution of (24). Now
\be
\tilde{\omega}_k=\omega_k(1-2/{\eta^2\omega_k^2})^{1/2}
\ee
As $\omega_k(\eta)$ is always real, we thus get
\be
\omega_k^2(\eta)=|\omega_k(\eta)|^2\simeq|\tilde{\omega}_k|^2(1+2/{\eta^2\omega_k^2})
\ee
Using this in (23) we get $k^2\eta^2+\alpha>2$. However this is not so important, it is mentioned here to show that $\omega_k^2$ is positive.Since
\bed
u_k=\frac{1}{\sqrt{2\tilde\omega}}\,e^{-i\,\int\tilde\omega_k d\eta}\;,\;\;\partial_{\eta}u_k=\frac{\tilde\omega_k^{\prime}}{2\,\tilde\omega_k}u_k-i\,\tilde\omega_ku_k\;\,{\rm i.e.}\;\;\left(\partial_{\eta}u_ku_k^*+u_k\partial_{\eta}u_k^*\right)=-\frac{\tilde\omega_k^{\prime}}{\tilde\omega}|u_k|^2
\eed
From the definition of $\tilde\omega_k$ we get $\tilde\omega_k^{\prime}=-\frac{\alpha-2}{\eta^3\tilde\omega_k}\;$ thus,
\bed
\frac{1}{\eta\omega_k^2}\left(\partial_{\eta}u_ku_k^*+u_k\partial_{\eta}u_k^*\right)=\frac{1}{\eta^2\omega_k^2}|u_k|^2
\eed
We can write
\beq
\frac{|u_k^{\prime}|^2}{w_k^2}&=&\frac{|u_k^{\prime}|^2}{|\tilde{\omega}_k^2|}+\frac{|u_k^{\prime}|^2}{\omega_k^2}-\frac{|u_k^{\prime}|^2}{|\tilde{\omega}_k^2|}\nonumber\\&=&\frac{|u_k^{\prime}|^2}{|\tilde{\omega}_k^2|}-\frac{2}{\eta^2\omega_k^2|\tilde\omega_k|^2}\left[|\tilde\omega_k|^2+\frac{1}{4\eta^2}\right]|u_k^2|\nonumber
\eeq
For small $\eta$ and neglecting the last term we have
\bed
\frac{|u_k^{\prime}|^2}{w_k^2}=\frac{|u_k^{\prime}|^2}{|\tilde{\omega}_k^2|}-\frac{2}{\eta^2\omega^2}|u_k|^2
\eed
Thus   
\be
N_k(\eta)=\frac{1}{4}\frac{\vert u_k\vert^2+\frac{1}{|\tilde{\omega}_k|^2}\vert u_k^\prime\vert^2-\frac{2}{\eta^2{\omega}_k^2}\vert u_k\vert^2-\frac{1}{\eta}\frac{\tilde{\omega}_k^\prime}{\tilde{\omega}_k\omega_k^2}\vert u_k\vert^2+\frac{1}{\eta^2\omega_k^2}\vert u_k\vert^2}{\vert v_{\vec{k}}\vert^2}-\frac{1}{2}
\ee
At small $\eta$, the last three terms in the numerator cancel out almost exactly. For large $\eta$ the contribution of last three terms also can be neglected when compared with the first two terms. We fix the initial vacuum  at large $\eta_i$ by taking
\be
v_k(\eta)=\frac{1}{\sqrt{2\tilde{\omega}_k}(\eta_i)}e^{i\tilde{\omega}_k(\eta_i)\,\eta}
\ee
Thus we get
\be
N_k(\eta)= \vert F_k(\eta)\vert[\frac{1}{2\vert\tilde{\omega}_k(\eta)\vert}\vert u^\prime_{\vec{k}}\vert^2+\frac{\vert\tilde{\omega}_k(\eta)\vert}{2}\vert u_{\vec{k}}\vert^2]-\frac{1}{2}
\ee
where
\bec
$|F_k(\eta)|=\frac{|\tilde{\omega_k}(\eta_i)|}{|\tilde{\omega}_k(\eta)|}=\frac{|(1+(\alpha-2)/{(k\eta_i)^2})^{1/2}|}{|(1+{(\alpha-2)}/{(k\eta)^2})^{1/2}|}$
\eec
We note some characteristic features of our expression (36).\\
(i) It is well known that WKB approximation is valid for large $\eta$. We show that WKB approximation as carried out in this paper, works very well in small $\eta$ region. In other words our expression (36) is valid for small and also for large $\eta$. So it is worthwhile to study these regions analytically. For large $\eta$, the prefactor $|F_k(\eta)|\rightarrow 1$ since the vacuum is defined at $\eta_i\rightarrow -\infty$. We then get back the result (30).\\
(ii) In WKB modes to define the vacuua are well known, usually known as Parker's definition. These modes serve as good template for particle and antiparticle states. Not only that, our definition used in the present work is also being currently used in the literature \cite{hmpm:grqc}. However this type of WKB definition has also relevance when one discusses squeezing to understand decoherence mechanism.\\
(iii) At $\eta\rightarrow 0$, the WKB mode solution is not valid. Here there is a standard analytic continuation to keep the WKB mode even at $\eta\rightarrow 0$. We first consider $m=0$ case in which analytical results are available for comparison. We get going to the variable $z=-k\eta$, the equation (6)as
\be
u^{\prime\prime}_k(z)+\tilde{\omega}^2_k(z)u_k=0
\ee
\be
{\rm and}{\hspace*{3cm}} N_k(z)=\frac{1}{2}|F_k(z)|\left[|\tilde\omega_k(z)||u_k|^2+\frac{1}{|\tilde\omega_k(z)|}|u_k^\prime(z)|^2\right]-\frac{1}{2}
\ee          
where now the prime denotes the differentiation with respect to $z$. Introducing a parameter $\lambda$ such that $\lambda^2\tilde{\beta}=\beta$ we write (37) as
\be
u^{\prime\prime}_k(z)-\lambda^2 \frac{\tilde{\beta}(z)}{z^2}u_k=0
\ee
where $\lambda$ is a large parameter and $\beta(z)=2-z^2$. The WKB solutions at $z\rightarrow 0$ is now given by
\be
u_k\sim e^{\pm\mu S}\sim z^{\frac{1}{2}\pm\sqrt{\beta}}
\ee
where
\beq
S&=&\int (2-z^2)^{1/2}\frac{dz}{z}\\\mu & = &\sqrt{1+\frac{1}{4\beta(0)}}
\eeq
The validity of WKB solution (40) can now be verified through the exact solutions of (39).
Doing exact solution to (39) we will find
\bec
$u_k=z^{\frac{1}{2}\pm\sqrt\beta\sqrt{1+\frac{1}{4\beta}}}$
\eec
This gives us (40), keeping WKB mode solution even at $z\rightarrow0$. With this change and evaluating 
\bec
$S(z)=\int^z\,\tilde{\omega}_k\,dz$
\eec
 we find after setting $\lambda^2\tilde{\beta}=\beta$
\be
N_k(z\rightarrow 0)\sim z \frac{1}{z^3}\sim \frac{1}{z^2}
\ee
where the extra factor $z$ comes from the $F_k(z)$ term. Without the $|F_k(z)|$ term, we will get $N_k(z)\sim 1/z^{2\sqrt{2}}$. If we do the exact calculations finding $u_k(z)$ in terms of Hankel functions we will get eqn (43). This is a very important result.
\par
So far we have nothing remarkable new. If particle production in de Sitter spacetime has anything to do with decoherence, we must be able to reproduce thermal spectrum at large $z$ and and large $N_k$ at small $z$. We know that the existence of horizon associates with the de Sitter spacetime a  temperature, which fact in our approach, will be taken into account through the introduction of complex trajectory. For $m^2/{2H_0^2}<<1$, well satisfied for vacuum fluctuation, we have negative $\tilde{\omega}_k^2$ at small $\eta$. Thus we have both growing and decaying modes, and if growing mode survives we have large $N_k$ and this allows us to connect $N_k$ to the existence of large squeeze parameter, a condition for having effective decoherence through particle production \cite{gjkks:ber}. In both the regions ($m^2>2H_0^2$ and $m^2<2H_0^2$) we will find that the complex trajectory plays the crucial role. Let us first discuss the emergence of thermal spectrum.    
\par
The emergence of thermal spectrum is straightforward in CWKB approach. We now write the expression of $N_k$ in terms of $z$. We have
\be 
N_k(z)=\vert F_k(z)\vert\left[\frac{1}{2\vert\tilde{\omega}_k(z)\vert}\vert u_k\,^\prime(z)\vert^2+\frac{\vert\tilde{\omega}_k(z)\vert}{2}\vert u_k(z)\vert^2\right]-\frac{1}{2}
\ee
We consider the case $\alpha>>2$. In that case we have no real turning point. The turning points are now complex, given by $z_{1,2}=\pm i \frac{m}{H_0}$. Keeping aside the pre-exponential factor in the WKB mode solution, we have
\be
u_k\sim e^{+i\int_{z_i}^{z}\tilde{\omega}_k(z)}
\ee
\be
u_k^\prime=i\tilde\omega_k(z)\,u_k
\ee
Now time flows from large $z$ to small $z$. Let us see how $u_2$ is modified when we take the contribution of complex turning point. Neglecting the vacuum contribution we have approximately (neglecting the variation of $\tilde{\omega}_k(z)$) using eqn (44)
\be
N_k(z)\simeq \frac{1}{2}\vert e^{i\int_{z_i}^{z}\tilde{\omega}_k(z)dz}\vert^2
\ee
Here $z_i$ is an arbitrary initial real point from where the trajectory starts and ends at the real point $z$ such that $z>z_i$ i.e., we are moving backward in time.  Now the trajectory starts from $z_i$, comes to a point $z_0$ (real) at which the particle becomes nonrelativistic, proceeds towards the complex turning point $z_1$, being reflected again comes to $z_0$ on the real axis and finally goes to $z$. It should be noted here that we are considering $u_k\simeq exp(iS)$ where $S$ is the WKB action (not to be confused with the scale factor). We write this contribution in steps 
\be
u_k\sim e^{iS(z_0,z_i)+2iS(z_1,z_0)+iS(z,z_0)}=e^{iS(z,z_i)+2iS(z_1,z_0)}
\ee
with  $S(z_f,z_i)$ being the WKB action between $z_i$ to $z_f$. The integral has been evaluated in \cite{bis2:cqg} to find
\be
N_k(z)\sim e^{-4Im\,\int_{z_0}^{z_2}(1+\frac{\alpha}{z^2})^{1/2}dz }=e^{-4\times\frac{\pi\,\alpha}{2}}
\ee
so that
\be
N_k(z)\sim e^{-\frac{m}{T}}
\ee
where $T=H_0/{2\pi}$ with Boltzmann constant taken as unity. Thus we get the thermalized spectrum. The point $z_0$ on the real axis, at which the trajectory starts towards the reflection,  can be found as follows. Starting with the scale factor $S(t)=H_0^{-1}exp(H_0t)$ and setting $S(t)=\pm\,i\frac{k}{m}$, we find the turning points $t_{1,2}=\pm\,i\frac{\pi}{2H_0}+H_0^{-1}ln\frac{kH_0}{m}$. At $Re\,t_{1,2}$ we get $a(Re\,t_{1,2})=\frac{k}{m}\equiv a(Re\,\eta)$, where particle becomes nonrelativistic. We will discuss this case again when we consider the case $m^2>>2H_0^2$ separately. One important feature is that the expression of $N_k(z)$ is nowhere is dependent on initial $z_i$ where the vacuum is defined. This is sometimes referred to high squeezing in momentum i.e., all differences in the initial state are damped out by inflation. There is a general result to evaluate the thermal spectrum for any expanding FRW spacetime. The result is 
\cite{sap:grg}
\be
\int_{z_0}^{z_1}\tilde{\omega}_kdz\simeq \frac{1}{(da/{dz})}_{z_0}\int_{a_0}^{a_1}\tilde{\omega}_k\,da=i\pi\,m/H_0
\ee
The temperature comes half of that obtained through the exact treatment. Though all FRW expanding spacetimes show thermal spectrum, large particle production is not found in all such cases except for the de Sitter case. Note that in evaluating (48) we consider only reflected trajectory contribution but there may be another direct trajectory from $z_i$ to $z$ to interfere with reflected part and will generate oscillation in $N_k$. We will discuss this separately.   
\par
We will now discuss the CWKB results of  particle production both analytically and through numerical calculation. We already noted that the expression of $N_k(z)$ does not depend on the initial moment $z_i$ but it correctly reproduces the vacuum condition $N_k(\infty)=0$ in the limit $z\rightarrow z_i\rightarrow \infty$. The regime of interest is $z\rightarrow 0$, when there is large expansion and the wavelength of the modes exceeds the Hubble radius. Here we are to consider two cases \\
{\bf (i) $m^2>>2H_0^2$}, real frequency with bounded classical motion and\\
{\bf(ii) $m^2<<2H_0^2$}, imaginary frequency and rolling down the hill.
\par
In the case of $m^2>2H_0^2$ the frequency is everywhere real and there is no turning point for real $z$. However there are complex turning points at $z=\pm iM$ with $M=\sqrt{m^2/{H_0^2}-2}$. It will generate a thermal spectrum with $N\simeq exp(-m/T)$ with $T=H_0/{2\pi}$ as already shown in eqn (50). However for quantum fluctuation, we are interested in those $m-$values for which frequency turns imaginary for $m^2/{H_0^2}<<\,2$. In this case the frequency turns imaginary at $z_0=i\sqrt{2}\sqrt{1-m^2/{(2H_0^2)}}$. If $\eta_0$ denotes the time when frequency becomes zero and $\eta_c$ the time at which the fluctuation crosses the Hubble radius, we get
\be
S(\eta_0)=S(\eta_c) \left(2-\frac{m^2}{H_0^2}\right)^{-1/2}
\ee
This relation implies that for the scalar field (supposed to have low mass $m^2<<H_0^2$) that drives the inflation, the one mode frequency turns imaginary before the mode crosses the Hubble radius. For $H_0^2<m^2<2H_0^2$, the ``rolling down the hill'' occurs even at later times, well after the physical wavelength of the mode crosses the Hubble radius. In other words, it is the evolution for $z<1$ that decides the fate of the fluctuation. We will discuss the case $m^2>2H_0^2$ later as it involves the inclusion of complex trajectory. \\
\\
{\bf Case :} $m^2<2H_0^2$
\\
 For the case $m^2<2H_0^2$ we are to evaluate $N(z)$ within the real turning points $\pm z_0$. So it is the evolution for $z<<z_0$ that decides the fate of the fluctuation. Here again we will apply the CWKB method to keep the particle-antiparticle definition fixed with respect to initial vacuum. From (38) we get the expression
\be
N_k(z)=\frac{1}{2}\left(e^{2Im\int\,\tilde{\omega}_k(z)dz}\vert F_k(z)\vert-1\right)
\ee
We now evaluate the integral
\be
I(z)=\int^z\,\tilde{\omega}_k(z)dz=Mlog\left(\sqrt{(z^2+M^2)}-M\right)-Mlog\,z+\sqrt{z^2+M^2}
\ee
where $M^2=(2-\frac{m^2}{H_0^2})$. With $m^2/{H_0^2}<<2$, taking the imaginary part of $I$ we  finally get
\be
N_0(z)=\frac{1}{2}\left[\vert F_k(z)\vert\left[\frac{z^2-2\alpha+4}{z^2}\right]^{\sqrt{(2-\alpha)}}-1\right]
\ee
where $\alpha = m^2/{H_0^2}$. It is evident from the expression that at $z\rightarrow \infty$, $N(\infty)=0$ satisfying the initial vacuum condition. In evaluating the above expression of $N_0(z)$ we have not taken the variation of frequency i.e., we have taken 
\be
u_k^\prime=i\tilde{\omega}_k(\eta)u_k
\ee
This $N_0(z)$ is plotted in fig. 3 for various values of $m/H_0$ and the production of large number of particles is quite evident. Let us see how this $N(z)$ is modified when we take particle production due to change in frequency.  We now get
\be
N_k(z)=N(z)=\frac{1}{2}\vert F_k(z)\vert\left[\left[\frac{z^2-2\alpha+4}{z^2}\right]^{\sqrt{(2-\alpha)}}\right]\left(1+\frac{1}{8\vert \omega_k(z)\vert^2}\left|\frac{\omega^\prime}{\omega_k}\right|^2\right)-\frac{1}{2}
\ee
This $N_k(z)$ is plotted in fig. 4. This fig. 4 is in in conformity with the variation of frequency shown in the fig. 2.  We obtain the fig. 2 using the expression
\bec
$\frac{\vert \omega_k^\prime\vert}{\vert \omega_k \vert}=\frac{\alpha-2}{\vert z \vert (\alpha-2+z^2)}$
\eec
where $\alpha=\frac{m^2}{H_0^2}$. 
\\
\\
{\bf{Case:} $m^2>2H_0^2$}\\
If we consider $\alpha>2$, the frequency is real for all real values of $z$, so from eqn (36)
\be
\left|u_k\right|^2\sim\frac{1}{2\tilde\omega_k(z)}
\ee
and hence in this case the particle production is entirely due to rapid change of frequency at small $z$. If we do not consider the turning points, which will be complex, the contribution like the first bracketed term in (57) will be now absent and hence we are to consider CWKB trajectory contribution more carefully. The importance of our approach is that the single expression for $N(z)$ represents all the characteristic features of particle production using only WKB approximation contrary to various approximation as is carried out in \cite{mm:prd}. 
\par
In this case the integral of $\omega_k(z)$ is always real and the particle number produced is much smaller compared to case $m^2<2H_0^2$.  If we take the contribution of the complex turning point at $\pm iM$ with $ M=\sqrt{\frac{m^2}{H_0^2}-2}$ we find that the $N_k(z)$ so reproduced does not result in any of the graphs shown in the figure (see \cite{mm:prd}, fig. 3). They obtained the graph with the Hankel function approximated to small $z$ values. Let us see what happens when we take complex trajectory approach. In this case there is no turning point at real $z$. The contribution of $u_k$ may be considered as
\be
u_k(z)=\frac{1}{\sqrt{2\tilde{\omega_k}(z)}}e^{i\int_{z_i}^z\tilde{\omega}_k(z)dz} 
\ee
where $z_i$ is the point where the initial vacuum is defined.The physical picture behind the CWKB particle production is Klein-paradox like situation , not in space but in time. Though this has been discussed elsewhere in details, we briefly discuss the physical picture to consider reflection in time as a process of particle production from vacuum. Consider the pair production from a time dependent potential $V(t)$ as
\bec
$V(t) \longrightarrow (e^+\uparrow)_+\,+\,(e^-\uparrow)_+$ 
\eec
Here the uparrow denotes that both particles move forward in time with positive energy. Now according to Feynmann-Stuckleburg prescription a positive energy particle moving forward in time is equivalent to negative energy antiparticle moving backward in time. Thus the pair production process shown above can be written as
\bec
$V(t) \longrightarrow (e^+\uparrow)_+\,+\,(e^+\downarrow)_-$
\eec
This equation implies that a reflection in time can be viewed as pair production \cite{bis1:pmn}. This corresponds to a situation where referred to eqn (24), particle production occurs at $\tilde{\omega_k}(z)=0$ The turning points may be real or complex. For real turning points we have situation just like tunneling through a potential barrier (case $m^2<2H_0^2$); for complex turning points, the situation we encounter is like over the barrier reflection in potential problems (case $m^2>2H_0^2$). The reader may consult \cite{bis3:grg,bcm:grg} to view the CWKB trajectories in tunneling and over the barrier situation in potential problems. In CWKB we consider 
(24) like Schr\"{o}dinger equation not in space but in time. The wave function at any real point is written as contribution of WKB trajectories. In CWKB, the variable (here $z$) is treated as complex. The wavefunction at any real point can now be written as sum of real trajectory and complex trajectory contributions. The incident, reflected and transmitted trajectories are then considered in the standard way. For details the readers may consult \cite{bis3:grg}. In de Sitter case, we have $z$ as the time variable. Thus $z\rightarrow -\infty$ correspond to the situation where there is no particle and $z\rightarrow +\infty$ correspond to the situation where we have pair production. The turning points in between the two regions given by $\omega(z)=0$ specify the barrier region. Here $exp[i\,S(z,z_i)]$ acts as incident wave and as it emerges from the complex turning point region, the wave is written as in eqn (66).
\par Previously while discussing the thermal spectrum we consider only the contribution of branch point of $\omega_k(z)$ and noted that the result is independent of $Re\,\,t$ on the real axis from where the trajectory starts  to get reflected. It is therefore necessary to consider the trajectories that go direct from $z_i$ to $z$ and also the trajectories that suffer reflection at the turning point and go to $z$. The trajectory that goes to $z_i\rightarrow z$ without any contribution from the turning point is
\bec
$e^{i\int_{z_i}^z\tilde{\omega}_k(z)dz}$
\eec
The other trajectory starting from $z_i$ goes to the turning point $z_1$ and turning back from $z_1$ again comes to $z$. This trajectory contribution is
\bec
$(-i)e^{i\int_{z_i}^z[\tilde{\omega}_k]dz}\,\,e^{2i\int_z^{z_1}\tilde{\omega_k}dz}$
\eec
 Here $(-i)$ has been introduced for reflection at $z_1$ and $\delta=\int^z\tilde{\omega}(z)dz=real$. We now introduce repeated reflections between the two turning points. Eqn.(59) is modified as \cite{bis3:grg}
\be
u_k^c=u_k(z)(1-i|R_c|e^{-2i\delta})
\ee
where
\be
|R_c|=\frac{exp(-\pi M)}{\sqrt{(1+exp(-2\pi M))}}
\ee
where the numerator comes from the repeated reflections between the turning points.
Hence we get
\be
|u_k^{(c)}(z)|^2=|u_k(z)|^2\left(1+\vert R_c\vert^2-2\vert R_c\vert \sin\,2\,\delta\right)
\ee
Now we find oscillation in $N_k(z)$. For $\alpha>2$, we have now the result (since $\delta$ is real)
\be
N_k(z)=\frac{1}{2}\left(1+\frac{1}{8\vert \omega(z)\vert^2}\left|\frac{\omega^\prime}{\omega}\right|^2\right)\left(1+\vert R_c\vert^2-2\vert R_c\vert \sin\,2\,\delta\right)-\frac{1}{2}
\ee 
In evaluating (63), we have not used the variation of $\delta$ with respect to $z$. If this variation is included, (63) is modified as
\beq
N_k(z)\equiv N(z) &=& \frac{1}{4}(1+R_c^2-2R_c\sin\,2\,\delta)+\frac{1}{4}|[-\frac{1}{2}\frac{\omega^\prime}{\omega}(1-R_c\sin\,2\,\delta)-\omega R_c\cos\,2\,\delta)]|^2\nonumber\\
     & & \,\,\,\,+|[\frac{1}{2} \frac{\omega^\prime}{\omega}R_c\cos\,2\,\delta+\omega+\omega R_c\sin\,2\,\delta]|^2
\eeq
We plot this $N(z)$ in fig. 5 for various values of $\alpha>2$. Interestingly, (64) is of the form
\be
N(z)=A_0(z)+A_c\,\cos\,2\,\delta+A_s\,\sin\,2\,\delta
\ee
where  $A_0 = -\frac{1}{2}+\frac{1}{4}(1\,+\,R^2)(2+\frac{1}{4|\omega|^2}|\frac{\omega^\prime}{\omega}|^2)$\\
$~~~~~~~~~A_c =+\frac{1}{2}\frac{\omega^\prime}{|\omega|^2}\,R$\\
$~~~~~~~~~A_s =-\frac{1}{8}\frac{1}{|\omega|^2}|\frac{\omega^\prime}{\omega}|^2\,R$\\
\\
Eqn (65) is of the same form as in \cite{mm:prd}. However our coefficients are different because in that work $z$ dependence in $\delta$ is introduced  in a different way. However we need not require any approximation for small $z$. The importance of our approach is that we need not require to evaluate the mode solutions and hence can be generalized to more complex cases. The present work is a good example in the cosmology where the CWKB paths play a definite role with a transparent physics. It should be noted that we worked always within WKB approximation and do not require to find the exact solution of (24) in terms of Hankel functions.
\par
In CWKB \cite{bis3:grg} $u_2$ for $z>0$ is written as
\be
u_2=e^{iS(z,z_i)}-iRe^{-iS(z,z_i)}
\ee
where $e^{iS(z,z_i)}$ is the incident trajectory and $e^{-iS(z,z_i)}$ is the reflected trajectory. Here we have not included the WKB pre-exponential factor for convenience.
For $m^2/H_0^2 < 2,$ the reflected trajectory contribution is very negligible. $e^{iS(z,z_i)}$ grows due to imaginary $\tilde\omega_k(z).$ For $m/H_0^2 > 2$ it is easy to show from the expression (66) that (62) follows with $R=R_c$ with the vacuum chosen as
\bec
$v_k=\frac{1}{\sqrt{\tilde\omega_k(z)}}e^{i\theta}$
\eec
Thus we see that the CWKB approach explains very successfully the cases $\alpha > 2$ and $\alpha <2$.
\section{Discussion of numerical results}
The numerical results are plotted in figs.1 - 5. In fig. 1 we plot frequency square for various values of the parameter $\alpha=m^2/{H_0^2}$. It shows that for $\alpha>2$ the frequency is unity at early times and changes rapidly as soon as the physical wavelength of the mode crosses the Hubble radius. The oscillator remains upright all the times. For $\alpha<2$ the oscillator turns to an upside down in the same regime. We note from (13) that there will be a term to particle production proportional to $\frac{\vert \omega^\prime_k\vert}{\vert \omega_k\vert}$. We plot it in fig. 2 in order to compare $N(z)$ for the cases $\tilde{\omega_k}^\prime=0$ and $\tilde{\omega_k}^\prime\neq 0$. To compare the contribution of changing frequency we plot in fig. 3 $N(z)$ with $u_k^\prime= i\tilde{\omega_k} u_k$  for various values of $\alpha\equiv\frac{m^2}{H_0^2}<2$. With the same value of $\alpha$ we plot $N(z)$ in fig. 4 taking into account the frequency change while evaluating $u_k^\prime$.
\par
 For $\alpha<2$, the frequency turns imaginary for $z<<1$. Here the CWKB paths take crucial role. 
\\
\begin{figure}[h]
\begin{centering}
\epsfig{figure=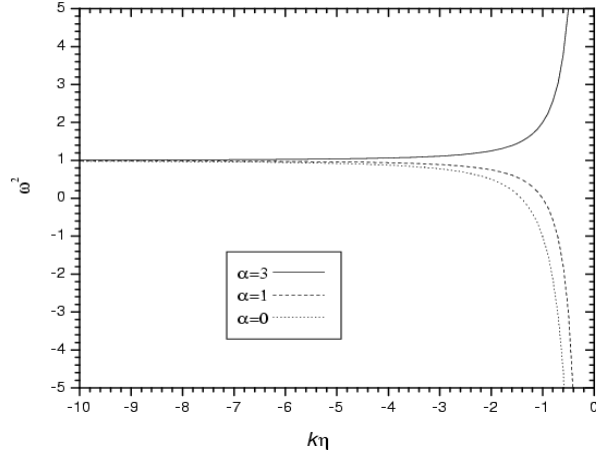,width=10cm,height=7cm}
\caption{The evolution of $\omega^2$ for various values of $\alpha=m^2/{H_0^2}$. For $\alpha$ greater or less than $2\,,\omega^2$ diverges after the physical wavelength of the mode crosses the Hubble radius. In the regime of large acceleration $0<\vert k\eta \vert<1$, the effect of changing frequency will be important.}
\end{centering}
\end{figure}

\begin{figure}[h]
\begin{centering}
\epsfig{figure=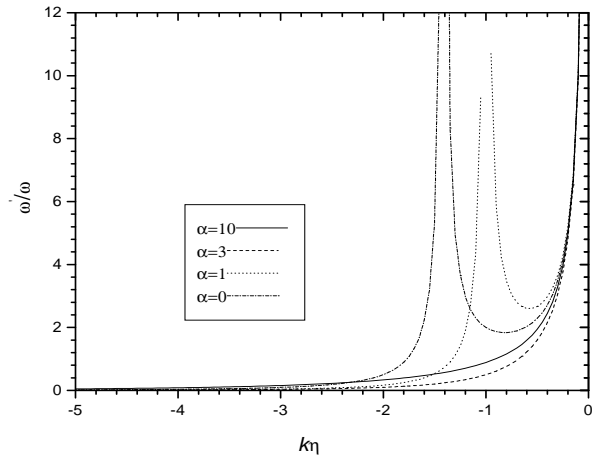,width=10cm,height=7cm}
\caption{Variation of $\frac{\omega^\prime}{\omega}$ with $k\eta$. The peaks in the figure come from the zero of $\omega^2$. For all values of mass the value of $\frac{\omega^\prime}{\omega}$ exceeds unity shortly after the physical wavelength crosses the Hubble radius.}
\end{centering}
\end{figure}
\vskip 1cm
\begin{figure}[h]
\begin{centering}
\epsfig{figure=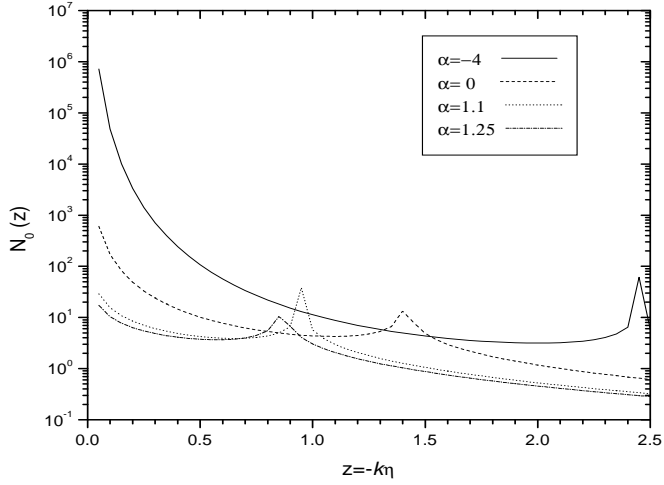,width=12cm,height=8cm}
\caption{The evolution of one mode occupation number in de Sitter space for different values of mass when the effect of $\frac{\omega^\prime}{\omega}$ is neglected.}
\end{centering}
\end{figure}
\vspace*{1cm}
\begin{figure}[h]
\begin{centering}
\epsfig{figure=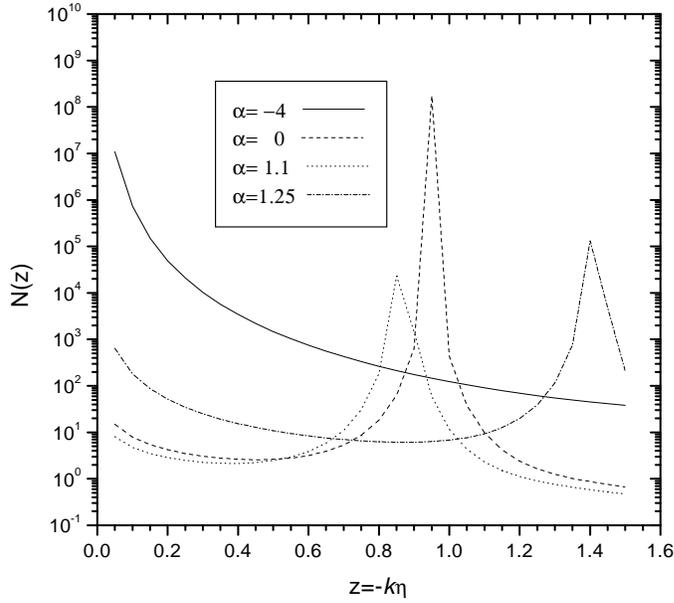,width=12cm,height=9cm}
\caption{The evolution of one mode occupation number in de Sitter space with the effect of $\frac{\omega^\prime}{\omega}$ included for $m^2/H_0^2 < 2$. The figure is obtained from eqn (57). The peak at $z=\sqrt2$ for $\alpha=0$ is due to $\omega^2$ passing through zero as the oscillator turns into a upside down. The corresponding divergence for $\alpha=-4$ is not shown in the figure as it is outside the range. For $\alpha=1.25$, we also get the peak at the correct position using the same eqn (57). The effect of $\frac{\omega^\prime}{\omega}$ is now clear when we compare fig. 4 with fig. 2 and fig. 3.}
\end{centering}
\end{figure}

\par
{\bf Case$\,\,\alpha<2$ :} Here we take $\alpha= -4, 1.25, 1.1, 0$ and plot the results in fig. 3 and fig. 4. The behaviour of $N(z)$ for the cases $\alpha= -4, 1.1, 0$ is the same as in \cite{mm:prd}, the peaks also comes in the right position in conformity with fig. 1. In our case the magnitude of $N(z)$ is larger in all the three cases and is obtained with CWKB approximation unlike the results in \cite{mm:prd}. This lends support to the fact our result for the case $\alpha=1.25\,\, \rm i.e.,\,\nu=1 $ is quite correct. Miji\'c \cite{mm:prd} treats this case as special case. The peak at $z=0.85$ corresponds to the zero of $\omega_k^2$ at this point.  The difference that we observe may be ascribed to the following facts. When we calculate $|\beta_k|^2$ using CWKB we get the last term of $N(z)$ as in (30). Obviously the approximation made in \cite{mm:prd} for $u_k$ will not lead to a factor $-i$ resulting in the factor $\frac{1}{2}$. With our WKB choice of modes we get the factor $-i$ i.e., we  measure $a^\dagger\,a$ with respect to the in vacuum with zero particle content i.e., $a_i\vert\, 0>=0$. The production of large number of particles at $z<1$ is a clear indication that we will get here the classical condensate. Let us now discuss the role of complex paths in CWKB.
\medskip
\par
{\bf Case$\,\,\alpha>2$ :} In fig. 5 we plot $N_k(z)\equiv N(z)$for five cases. For $\alpha=6.25$ the oscillation is not transparent because of smallness of $N_k(z)$. We found the oscillations in all the graphs (plotted without logscale). We may now easily calculate the period of oscillation setting  $\delta=\pi$. In terms of $z$ this gives $z_1/z_2=exp[\pi/\sqrt{\nu^2-\frac{1}{4}}]$, almost the same as in \cite{mm:prd}. Another aspect of our result is that at small $z<<1$ i.e., for long-wavelength mode outside the Hubble radius the occupation number diverges as
\be
\vert N_k\vert \sim O(1)z^{-2\sqrt{2-\frac{m^2}{H_0^2}}}
\ee
using the expression of $N_k(z)$ for $\alpha<2$. Thus for physical wavelength $\lambda_{phy}=1/(H_0\,z)$ we get
\be
\vert N_k\vert\sim \left(\frac{\lambda_{phy}}{H_0^{-1}}\right)^{2\sqrt{\nu^2-1/4}}
\ee
This result also coincides with Miji\'c except the factor $1/4$. At $z\rightarrow 0$, we are to consider the analytic continuation as in (39). The emergence of this oscillation also follows from squeezed state formalism \cite{aa:hep}. We discuss this briefly in concluding section. 
\par
From the discussion of analytical and numerical results, we have the important results that for $\alpha < 2 $ we get particle production with large $N_k$ favouring the formation of classical condensate. In squeezed state approach to de Sitter spacetime one can show \cite{gjkks:ber}, $N_k=sinh^2 r_k$ where $r_k$ is called the squeeze parameter. A large $r_k$ is necessary to obtain effective decoherence.
\par
For $\alpha > 2$ the modes are well within the horizon as already discussed in this paper just after the eqn (52). It has also been shown by Albrecht et al \cite{aa:hep} that in such cases the oscillatory solution becomes reasonable approximation. This is what we observe in our CWKB calculation leading to an oscillating $N_k$.
\begin{figure}
\begin{centering}
\epsfig{figure=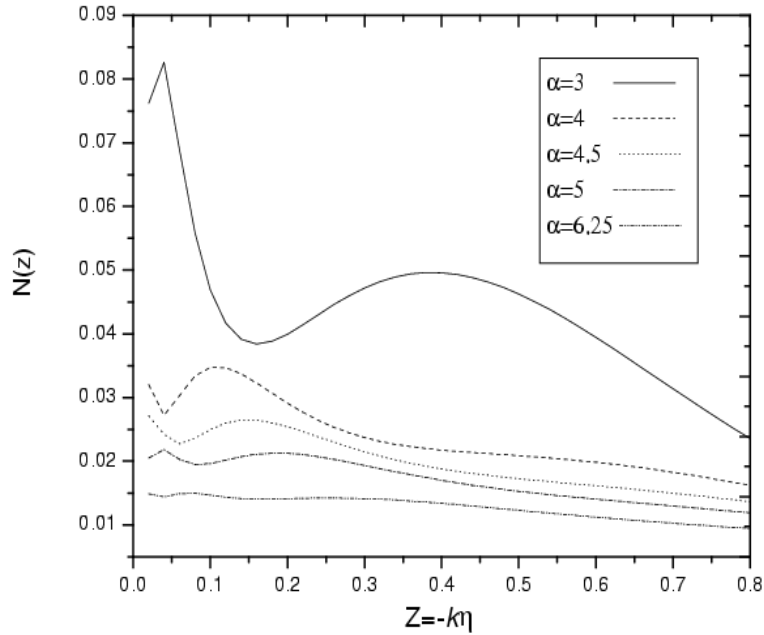,width=12cm,height=10cm}
\caption{The evolution of one mode occupation number in de Sitter space for different values of mass with $\alpha>2$. The figure is plotted with eqn (64). The oscillation is prominent. For $\alpha=25/4$ there are also oscillations, not prominent in the figure. Here the particle number is observed to be oscillatory function of $\mid{k\eta}\mid$; there is no formation of classical condensate.}
\end{centering}
\end{figure}
\section{Conclusion}
The calculation shows that in de Sitter spacetime the classical condensate can only form for the case $\alpha<2$ i.e., the inflation field is very weakly coupled and nearly massless. We derive all the essential results from a single formula using CWKB modes. We find that the parameter $m/H_0$ plays a vital role in the formation of classical condensate . The role of de Sitter horizon distance $H_0^{-1}$ and the Compton wavelength $m=\lambda^{-1}$ is now clear. Let us generalize the result by introducing a nonminimal coupling term $\xi R\phi^2$ whose effect is to replace $m^2$ by $m^2+12\xi H_0^2$ and $\alpha$ by $\sqrt{\frac{m^2+12\xi H_0^2}{H_0^2}-2}$. For $\alpha>2$, no classical condensate is formed and for sufficient large $z$ when oscillation dies out the occupation number behaves as Planckian spectrum independent of $z$ with 
\be
N=N_0\,\left(e^{-\frac{M}{T}}-1\right)^{-1}
\ee
where $M^2\equiv\frac{m^2+12\xi H_0^2}{H_0^2}>0 $ i.e., we recover the Planckian spectrum.  $M^2<0$ is the region for the formation of classical condensate. As most of the results obtained in this work substantiate the results obtained in \cite{mm:prd}, without going into further discussion, we let the readers read the concluding section of the cited work. What is new is the role of CWKB trajectory in generating the Planckian spectrum and discussion of decoherence within the squeezed state formalism. Our CWKB expression of particle production is so simple that we need not require to keep track the correctness of the expressions employed in each stage for various values of $m/H_0$ \cite{mm:prd}. This is the advantage of our approach. Moreover in order to clear up the criticism against the Miji\'c's works (eqns (26) and (27) in the present work), we use Hamiltonian diagonalization  method to obtain the expression of $N_k(z)$ valid for all $z$. This needs a minor correction to Miji\'c's calculation that we have shown without changing the basic conclusions of their work. Many feel discomfort with Hamiltonian diagonalization method. Our approach is an example where one can proceed with WKB definition of vacuum instead of Hamiltonian diagonalization. 
\par 
A common statement that is made while discussing decoherence is that decoherence is proportional to the strength of the coupling and it vanishes if the coupling goes to zero. The analysis carried out here for free modes shows that we still get decoherence without being dependent on coupling constant. The reason for such a appearance of coupling-independent decoherence may be due to the some specific combination of fluctuations and the inflaton zero mode that can assemble to generate a new effective field . It is this object which actually rolls down. This interpretation saves the standard picture of how metric perturbation are generated during inflation with the spinodial growth of fluctuations dominating the late time behaviour of Bardeen variables for superhorizon modes during inflation. This view is expressed in \cite{boy:atp}. 
The reason for studying the particle production through CWKB is twofold. This allows us to study parametric and spinodial resonances both in FRW and de Sitter spacetime under wide range of initial conditions. It should be pointed out that we find parametric resonances in chaotic inflationary scenarios for unbroken symmetry ($m^2 > 0$) whereas spinodial instabilities occur in new inflation scenario in broken symmetry ($m^2 <0$). As this is beyond the scope of the present work , we refer the reader to see the role of WKB modes in \cite{bis2:cqg} for a study of out of equilibrium fields in inflationary dynamics and its relation to density fluctuations. We like to point out a connecting feature of the present work to the standard approach of decoherence in which classicality is studied through the squeezed mode parameters. We show that the emergence of squeezed modes is a very natural outcome in the CWKB framework with $\omega_k$ having turning points at real or complex time, hitherto not discussed in the literature.  
\par
To relate our work to decoherence, we note that proceeding as in \cite{aa:hep}, one can easily show that the squeeze parameter $r_k$ becomes large because $N_k$ is large at $z\rightarrow 0$ for $m^2/{H_0^2} <2$ since $N_k=sin^2h\,r_k$. Not only that we have been able to generate oscillating $N_k$ for $m^2/{H_0^2} >2$ (also obtained in \cite{aa:hep}) along with the emergence of thermal spectrum with correct de Sitter temperature. In most of the earlier works dealing with squeeze state formalism (see \cite{gjkks:ber}) and in \cite{mm:prd} emphasis has been put on largeness of $r_k$ to understand decoherence in terms of large particle production. The emergence of oscillating $N_k$ and thermal spectrum have not been discussed effectively. In our view the emergence of thermal spectrum plus the large particle production are crucial to understand the emergence of classical world. In the present work we show that the CWKB approach reproduce these two aspects very clearly. Not only that all the conclusions that one gets through squeeze state formalism also follow in this approach which we termed as decoherence without decoherence.

\end{document}